\documentclass[iop]{emulateapj}
\usepackage{hyperref,url}
\usepackage{natbib,graphicx,amsmath}
\shorttitle{AO-assisted variability in globular clusters}
\shortauthors{Salinas et al.}
\journalinfo{To Appear in AJ}

\begin{document}

\title{An AO-assisted variability study of four globular clusters
  \footnotemark[$\dagger$]}

\author{R. Salinas$^{1,2}$, R. Contreras Ramos$^{3,4}$, J. Strader$^2$, \\P. Hakala$^5$, M. Catelan$^{3,4}$, M. B. Peacock$^2$ and M. Simunovic$^{4}$}
\email{rsalinas@gemini.edu}

\affil{$^1$ Gemini Observatory, Casilla 603, La Serena, Chile}

\affil{$^2$ Department of Physics and Astronomy, Michigan State
  University, East Lansing, MI 48824, USA} 

\affil{$^3$ Millennium Institute of Astrophysics, Av. Vicu\~na Mackenna 4860, 782-0436 Macul, Santiago, Chile}

\affil{$^4$ Instituto de Astrof{\'i}sica, Pontificia Universidad Cat\'olica de Chile, Av. Vicu\~na Mackenna 4860, 782-0436 Macul, Chile}

\affil{$^5$ Finnish Centre for Astronomy with ESO, University of
  Turku, V\"ais\"al\"antie 20, FI-21500 PIIKKI\"O, Finland}

\begin{abstract}
 
 The image subtraction technique applied to study variable stars in
 globular clusters represented a leap in the number of new detections,
 with the drawback that many of these new light curves could not be
 transformed to magnitudes due to the severe crowding. In this paper
 we present observations of four Galactic globular clusters, M\,2 (NGC
 7089), M\,10 (NGC 6254), M\,80 (NGC 6093) and NGC 1261, taken with
 the ground-layer adaptive optics module at the SOAR Telescope,
 SAM. We show that the higher image quality provided by SAM allows the
 calibration of the light curves of the great majority of the
 variables near the cores of these clusters as well as the detection
 of new variables even in clusters where image-subtraction searches
 were already conducted. We report the discovery of 15 new variables
 in M\,2 (12 RR Lyrae stars and 3 SX Phe stars), 12 new variables in
 M\,10 (11 SX Phe and one long-period variable) and one new W UMa-type
 variable in NGC 1261. No new detections are found in M\,80, but
 previous uncertain detections are confirmed and the corresponding
 light curves are calibrated into magnitudes. Additionally, based on
 the number of detected variables and new \textit{HST}/UVIS
 photometry, we revisit a previous suggestion that M\,80 may be the
 globular cluster with the richest population of blue stragglers in
 our Galaxy.
 
\end{abstract}


\section{Introduction} \label{intro}

\let\thefootnote\relax \footnotetext{$^{\mathrm{\dagger}}$Based on
  observations obtained at the Southern Astrophysical Research (SOAR)
  telescope, which is a joint project of the Minist\'{e}rio da
  Ci\^{e}ncia, Tecnologia, e Inova\c{c}\~{a}o (MCTI) da Rep\'{u}blica
  Federativa do Brasil, the U.S. National Optical Astronomy
  Observatory (NOAO), the University of North Carolina at Chapel Hill
  (UNC), and Michigan State University (MSU).}

The study of variable stars in crowded environments (e.g. globular
clusters) has a long discovery history that was boosted by the
introduction of image subtraction techniques
\citep[e.g.][]{tomaney96,alard98,alard00} that allow the discovery of
variable stars even in the cores of dense clusters. The standard
approach in this technique is to convolve a good seeing image, after
image registration, to match its point spread function (psf) to a
series of poorer seeing images, before the image subtraction which
leads to images where, in principle, only the sources that change
within the time scale of the observations, such as transients or
variable stars, are left.

Even though image subtraction has an impressive ability for the
\textit{detection} of new variables in crowded environments
\citep[e.g.][]{contreras05}, for many of these stars, crowding and
blending make it extremely difficult, if not outright impossible, to
transform the relative flux light curves provided by image subtraction
into magnitudes \citep[e.g.][]{baldacci05,corwin06,lazaro06}, often
leaving out of reach much of the information that could be obtained
from them (e.g. amplitudes, mean magnitudes, color-magnitude diagram
positions, and even information on cluster membership).

In this paper we explore the use of time--series photometry aided by
adaptive optics (AO) to obtain sharper images in four Galactic
globular clusters: M\,2 (NGC\,7089), M\,10 (NGC\,6254), M\,80
(NGC\,6093) and NGC 1261. Sharper images allow to obtain absolute
photometry of a larger number of stars in crowded environments and
therefore a calibration of the light curves into standard magnitudes.

We present the instruments used and the data obtained together with
its reduction in Section \ref{sec:obs}, while the detection of
variables is presented in Section \ref{sec:vars}. Periods and
classification for the new variables are given in Section
\ref{sec:variables}. Section \ref{sec:m80} presents a re-assessment of
the blue straggler (BS) content of the dense cluster M\,80, while a
summary is given in Section \ref{sec:conclusions}.

\section{Observations and data reduction} \label{sec:obs}

\subsection{SOAR/SAM: AO optical imaging}

\begin{deluxetable*}{lccccr}
\tablewidth{0pt}
\tablecaption{Observing log}
\tablehead{
\colhead{Cluster} &  \multicolumn{3}{c}{N$\times$Exptime} & \colhead{UT date} & \colhead{Time span}\\
&\colhead{$g$} & \colhead{$r$} & \colhead{$i$} & &\\
& (seconds) & (seconds) & (seconds)& (mm.dd.yyyy)& (hours)
}
\startdata
M\,2      & $24\times30$ & $3\times30$ & $228\times30$  & 07.21.2015 & 2.84\\
M\,10     & $20\times60$ & $2\times60$ & $278\times60$  & 07.21.2015 & 6.65\\
M\,80     & $3 \times60$ & $300\times60$ + 15$\times$80 & $3\times60$& 04.23.2014 & 5.98\\
NGC 1261 & $13\times90$ & $87\times60$ & --& 12.07.2014 & 2.36
\enddata
\label{table:log}
\end{deluxetable*}

Time series imaging of M\,2, M\,10, M\,80 and NGC 1261 were obtained
using the SOAR Adaptive Module (SAM) coupled with its Imager (SAMI),
installed at the SOAR 4.1m telescope at Cerro Pach\'on, Chile. SAM is
a ground layer adaptive optics system correcting atmospheric
turbulence near the ground. Technical details of the instrument can be
seen in \citet{tokovinin10,tokovinin12}. SAMI provides a field-of-view (FOV) of
$3\arcmin\times3\arcmin$ with a pixel scale of 0.0455\arcsec. Observations of
M\,2, M\,10 and M\,80 were taken with a $2\times2$ binning, while the NGC
1261 images were unbinned.

Table \ref{table:log} gives an observing log indicating dates,
filters, exposure times and the total time span of observations. Time
series imaging was conducted primarily using the SDSS $r$ and $i$
filters, selected as a compromise between the improvement in the image
quality given by SAM (better to redder wavelengths), the desire to
study, among others, pulsating \textit{blue} stragglers and the sky
brightness during gray time. Additional SDSS $g$ images were taken to
place the variables on color-magnitude diagrams and gain further
insight into their nature (Section \ref{sec:variables}).

The time span of the observations varied from cluster to cluster,
being at most six hours. This is inadequate for a good description of
the complete light curve of RR Lyrae (RRL) stars ($P\sim13$ hours for
ab pulsators and $\sim 7$ hours for c pulsators) commonly found in
metal-poor clusters, but sufficient for their detection and, albeit
uncertain, classification. For short-period variables like SX
Phoenicis-type stars (hereafter SX Phe), this time span covers about
2--6 complete pulsation cycles, sampling the complete light curve and
allowing good period determination. This dataset is inadequate for
long--period variables such as Miras or the brighter type II Cepheids,
including W~Vir and RV~Tau stars \citep[e.g.][]{catelan15}.

\begin{figure}
\includegraphics[width=0.48\textwidth]{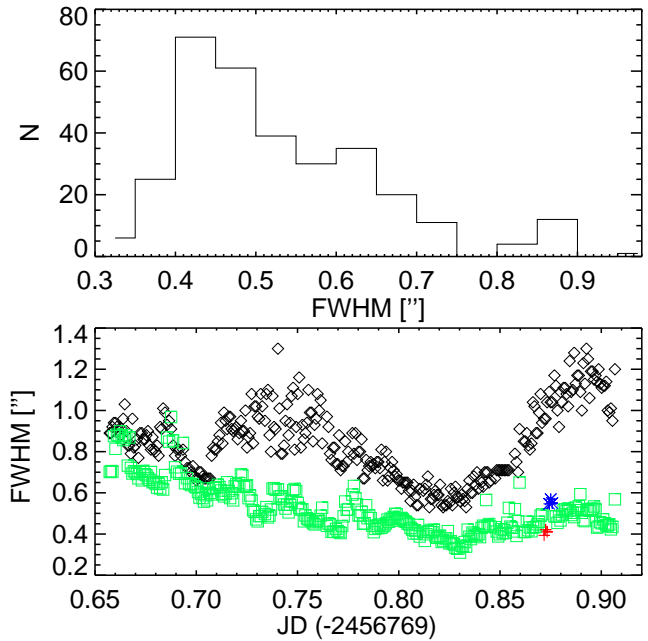}
\caption{SAM image quality. Top panel: A histogram of the measured 
FWHM on the complete SAM dataset of M\,80. Lower panel: The evolution of 
seeing as a function of time during the night. Black rhombi are 
the DIMM seeing measurements while green squares are the measured FWHM on the $r$ 
SAMI images. Red crosses represent the measurements on the $i$ images, 
while blue asterisks are the $g$ measurements.
\label{fig:seeing}}
\end{figure}

The image quality that can be achieved with SAM is exemplified in
Fig. \ref{fig:seeing} for the cluster M\,80. The lower panel shows the
variation of the natural seeing (black rhombi) and the average FWHM as
measured using the \textsc{gemseeing} task within IRAF for each image
(green squares for $r$ images, red for $i$ and blue for $g$), during
the time span of the M\,80 observations. Even though at the beginning
of the observations the improvement was only of the order of
0.1\arcsec, once the natural conditions dropped to around 0.6\arcsec\,
seeing, SAM provided 0.35\arcsec\, seeing images in the $r$
filter. During the rest of the night, the correction was of the order
of 50\%. The top panel shows a histogram of the measured FWHM in the
$r$ filter. The median FWHM was 0.49\arcsec.

Atmospheric conditions were significantly poorer during the other
dates. The median measured FWHM for M\,2, M\,10 and NGC 1261 in ($g$,
$r/i$), were (0.84\arcsec ,0.65\arcsec), (0.89\arcsec,0.69\arcsec),
and (0.75\arcsec ,0.68\arcsec), respectively.

The \citet{landolt92} standard field Mark-A was observed twice (in
open loop) in $gri$ to obtain a calibration into the standard system
during the M\,80 run. Magnitudes for these stars in the SDSS sytem
were provided by Elizabeth Wehner$^1$\footnote{$^1$
  \url{http://physwww.mcmaster.ca/~harris/sloan_standards.dat}}. Given
the relatively small FOV of SAMI compared to the Landolt fields, only
two Landolt stars were included and the calibration must be considered
as approximate. The derived transformation equations, obtained with
\textsc{iraf/photcal}, were:

\begin{align*}
g_{\rm instr}&=g_{\rm std}-1.063-0.065(g_{\rm std}-r_{\rm std})\\
r_{\rm instr}&=r_{\rm std}-0.922-0.027(g_{\rm std}-r_{\rm std})\\
i_{\rm instr}&=i_{\rm std}-0.767-0.057(r_{\rm std}-i_{\rm std}),
\end{align*}
where the rms of each fit is 0.04, 0.07, 0.04 mags respectively.

SOAR/SAMI data were reduced with the Python-based SAMI pipeline,
developed by Luciano Fraga at SOAR. The pipeline bias subtracts, flat
fields, mosaics the data read by the four SAMI amplifiers and provides
a rough astrometric solution. The astrometry was then refined using
\textsc{msctpeak/ccmap} in IRAF.

\begin{figure*}
\includegraphics[width=0.98\textwidth]{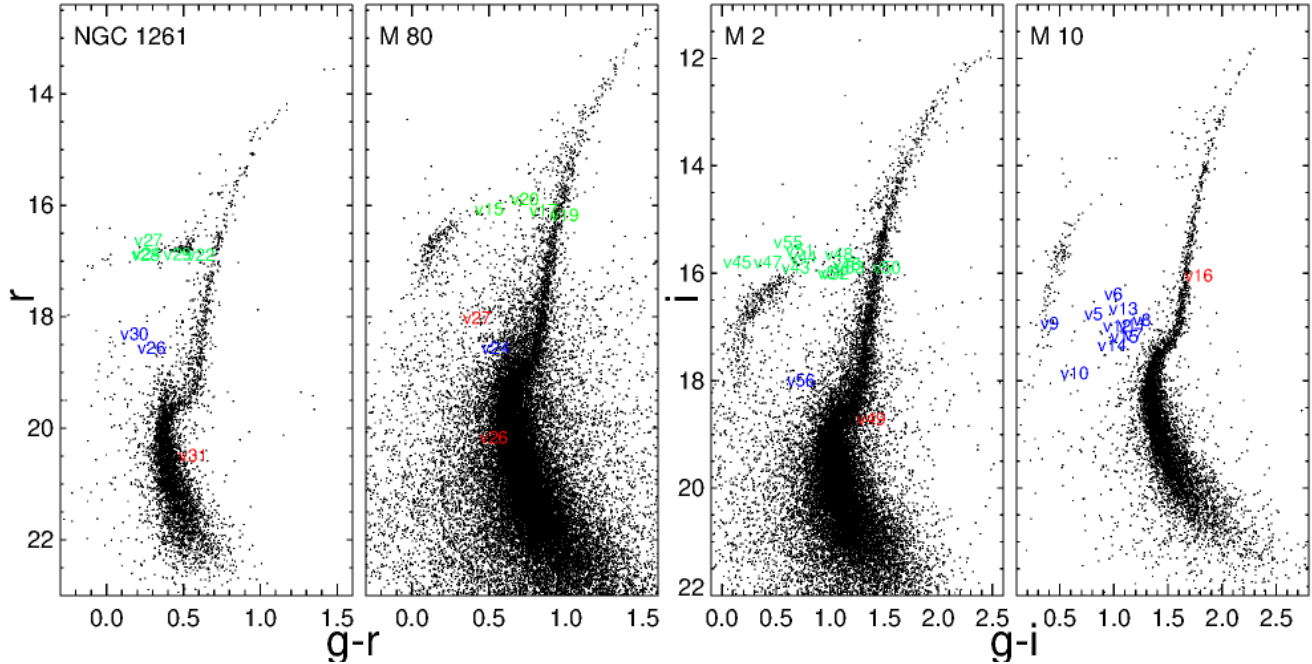}
\caption{CMDs obtained with \textsc{allframe} photometry for M\,2, M\,10, M\,80
  and NGC 1261 using SAM data. RRL stars are indicated in green, blue
  stragglers, in blue, while other variables (long period, eclipsing),
  in red.
\label{fig:cmd_sam}}
\end{figure*}

\subsection{HST/UVIS imaging}\label{sec:hst}

M\,80 is one of the densest clusters in our Galaxy, with the claim that
it hosts the largest amount of blue stragglers of any cluster
\citep{ferraro99}. As such, a large number of (eclipsing) binaries and
pulsating blue stragglers should likely be expected.

Given the high central density of M\,80, as an aid to the SAM imaging we also
retrieved WFC3/UVIS data from the \textit{HST} public archive 
that was collected in June 2012 (GO-12605, PI: Piotto) and consist of 
10 F255W (855 s), 5 F336W (657 s) and 5 F438W (85 s) exposures. The imaging strategy was optimized to minimize the well-known charge transfer efficiency problems of UVIS \citep[see][for details]{piotto15}.

Stellar photometry of the dark and bias subtracted and flat-fielded
data was performed using the online public program
\textsc{img2xym\_wfc3uv}, which is based on the photometry package
developed for the ACS/WFC camera \citep{anderson06}. This software was
specifically designed to perform reduction analysis of the
under-sampled WFC3 data using empirical psfs. Fluxes and positions
were corrected for geometric distortions using the solution given by
\citet{bellini11}. In order to place our catalogue in a common
coordinate system we have used astrometric standard stars selected
from the third US Naval Observatory CCD Astrograph Catalog
\citep[UCAC3,][]{zacharias10} and used \textsc{CataXcorr}, a program
developed at the Bologna Observatory (P. Montegriffo, private
communication), to perform roto-translation procedures. Our final
catalog consists of highly internally precise photometry and
astrometric positions of $\sim$40000 stars around the center of M80,
from the tip of the RGB to $\sim$4 magnitudes below the main sequence
turn off (see Section \ref{sec:m80}).

\section{Variable stars detection and photometry}\label{sec:vars}

\subsection{Image subtraction} \label{sec:isis}

Variable stars were initially searched using \textsc{isis} \citep[v
  2.1,][]{alard00}, an implementation of the image subtraction
technique which registers and matches the point spread function of
images before subtraction, over the SAM dataset. A reference frame
that is convolved to match the psf of each image was constructed using
a handful of images with the lowest FWHM.

Once the convolved reference frame is subtracted from each individual
image, ISIS produces a stack of all the absolute residual images where
variables can now be searched. We find that a visual inspection of the
stacked residuals produces the best results instead of applying a
detection threshold on this image \citep{salinas05,salinas07},
discarding false residuals produced by cosmic rays, bad pixels and
saturated stars.

As a further search for variables in the region of the blue
stragglers, we searched for variability in the positions of all the
blue stragglers as determined from the CMDs using SAMI data (see Fig.
\ref{fig:cmd_sam}) as well as the \textit{HST}/UVIS CMD for M\,80 (see
Section \ref{sec:m80}). Furthermore, BSs were also identified in the
\textsc{DAOphot} catalogues from UVIS observations of M\,2 and M\,10 retrieved
from the Hubble Legacy Archive$^2$\footnote{$^2$\url{hla.stsci.edu}}. For
NGC 1261 we used the \textit{HST} photometry from
\citet{simunovic14}. The success of this approach is exemplified in
NGC 2808, where \citet{catelan06} found a higher number of pulsating
blue stragglers than other studies which used the same
\citep{corwin04} or augmented \citep{kunder13} datasets, but different
variable identification approaches.

Relative flux light curves provided by ISIS were then turned into
magnitudes using the \textsc{allframe} photometry (see Section
\ref{sec:psf}), following the procedure outlined in
\citet{catelan13}. Calibration of the light curves into the standard
system formally requires color information for each epoch. Color
variations within a pulsation cycle, for example, for RRL stars, will
not be larger than about $V-R=0.3$ mag \citep[e.g.][]{kunder10}. In
practice assuming a constant $g-r$ (or $g-i$) color during the
calibration introduces a systematic error of $\leq0.02$ mag in $r$, a
value that is smaller than the uncertainty introduced by the
calibration equations (Sect. \ref{sec:obs}.

\begin{figure*}
\includegraphics[width=0.98\textwidth]{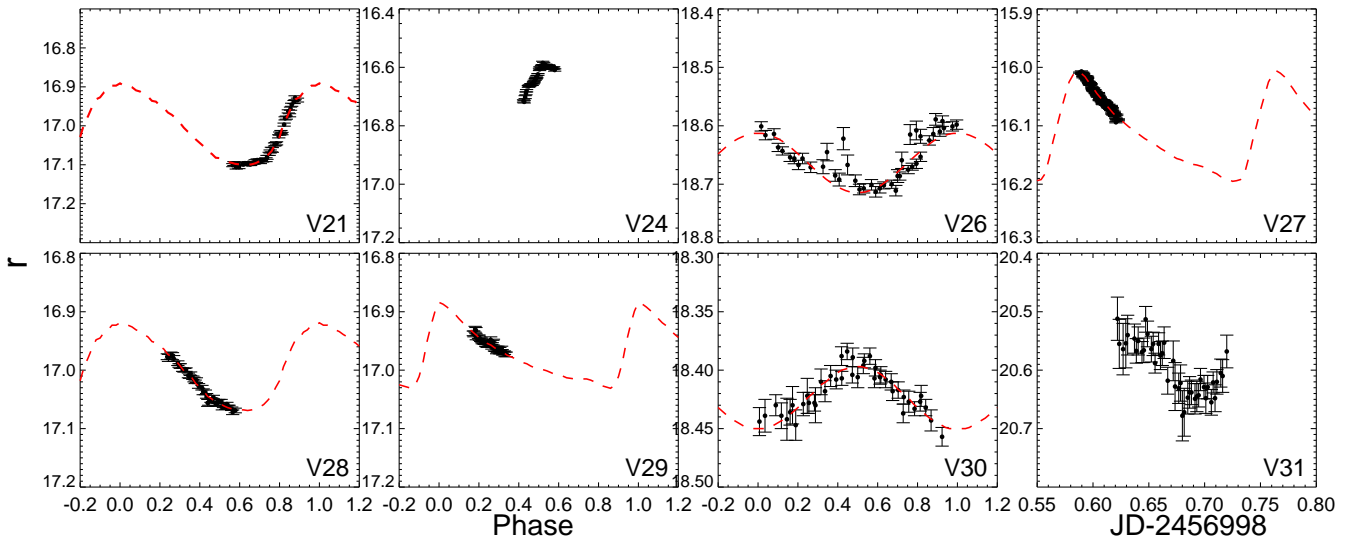}
\caption{Variables in NGC 1261. V21-V30 were discovered by
  \citet{salinas07}, while V31 is a new discovery. Dashed lines
  indicate the best-fit template. The V31 light curve is not shown
  phased, but only with its Julian date, given the large uncertainty
  in its period. \label{fig:n1261_lc}}
\end{figure*}

\subsection{PSF photometry} \label{sec:psf}

In order to construct color-magnitude diagrams that help pin down the
nature of the found variables and provide the reference magnitudes
used to transform the ISIS relative flux light curves into magnitudes,
stellar photometry was also conducted using the stand-alone
\textsc{daophot/allstar} profile-fitting photometry package
\citep{stetson87}. The psf was constructed by choosing between 50 and
100 bright and isolated stars, which were modeled as a quadratically
varying Moffat function (with $\beta=2.5$) following previous work
with SAM \citep{fraga13}. The same psf stars were used in all frames
for each cluster and their coordinates were transformed using
\textsc{daomatch/daomaster} \citep{stetson93} and \textsc{stilts}
\citep{stilts}.

Model psfs and \textsc{allstar} photometry were then used to obtain
improved photometry with \textsc{allframe} \citep{stetson94},
following the procedure described in \citet{salinas12}. Resulting
color-magnitude diagrams for the four clusters can be seen in
Fig. \ref{fig:cmd_sam}.

\section{Known and newly found variables}\label{sec:variables}

In this section we give details about the new variable stars detected
following the procedure presented in Section \ref{sec:vars}, as well as
quantities not given before for the previously known variables.

Periods for the new variables were determined using the phase
dispersion minimization method \citep{stellingwerf78} as implemented
in IRAF. No attempt was made to refine the periods for the known
variables given the short time span of the observations. Some SX Phe
variables are known to pulsate in more than one mode, generating
amplitude variations \citep{Eggen52,Walraven53}. Multiple periods in
our SX Phe candidates were searched using the standard Lomb-Scargle
technique \citep{scargle82}.

The limited phase coverage for RRL stars also precludes giving
reliable amplitudes and mean magnitudes directly from the data. These
quantities were estimated instead, when periods were known, by using
the template fitting code of \citet{layden98}. The code fits ten
templates of RRL stars and eclipsing binaries for a given period,
giving a $\chi^2$ value that is used to determine the best template,
and optionally, its classification. Being the longest period
variables, RRab are the most affected by the limited phase coverage
and their amplitudes are in many cases underestimated. Even though the
templates were set up using observations in $V$ \citep{layden98}, the
general shape of RRL remains mostly unaltered with passband, changing
only their amplitudes, which is taken into acount in the fitting
process.

\subsection{NGC 1261}

The variable star content of NGC 1261 was studied by \citet{wehlau77a}
and \citet{wehlau77b}, finding 18 RRL and one long period variable. A
modern image-subtraction search was conducted by \citet{salinas07} who
further found 4 RRL, 3 SX Phe variables and one more long-period
variable. \citet{salinas07} provided only relative-flux light curves
and periods, so here we give light curves in magnitudes, using the
limited phase coverage of our SAMI data, their position on a CMD, and
the estimated intensity-weighted mean magnitudes and amplitudes
resulting from the template fitting procedure in the case of the
RRL. No satisfactory was found for V24. For the SX Phe these are
obtained directly from the data (Table \ref{table:vars_n1261}).

\begin{figure*}
\includegraphics[width=0.98\textwidth]{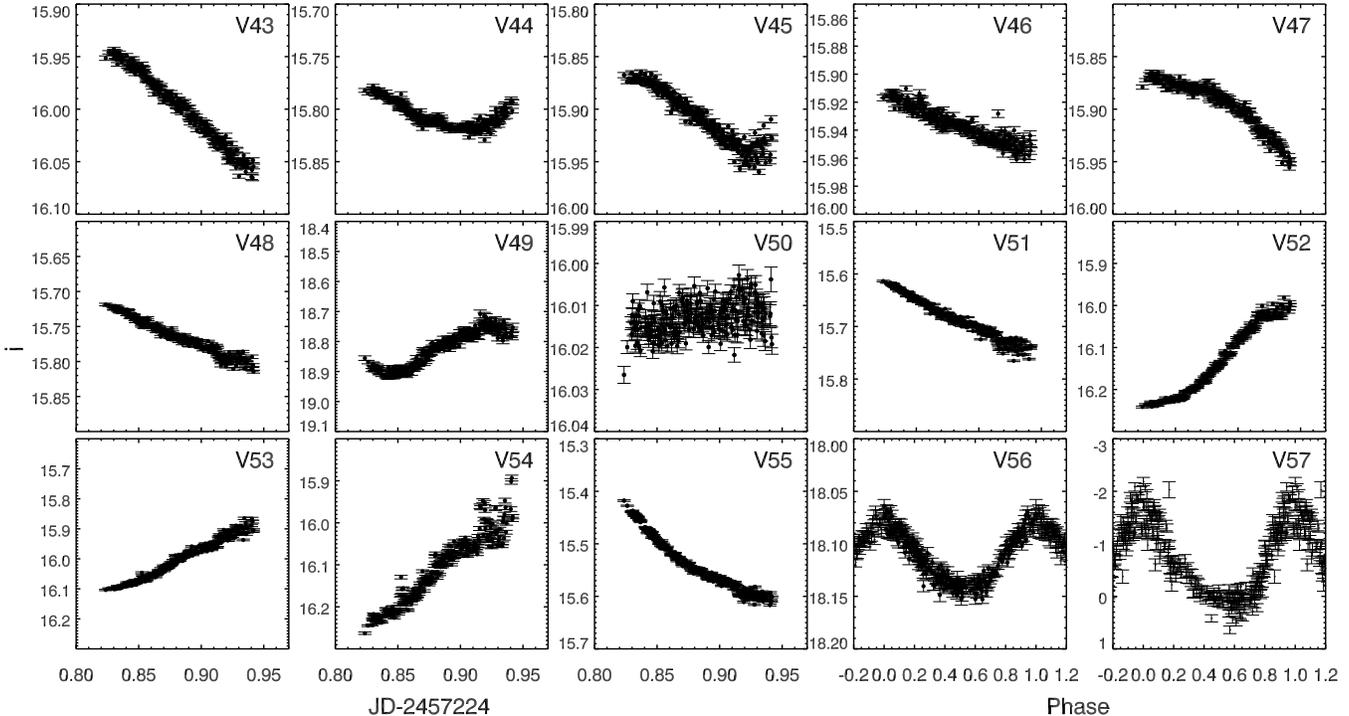}
\caption{New variables in M\,2. Variables V43 to V55 are shown in
  Julian dates due to our inability to find reliable periods based on
  the short time span of the observations. V56 is shown phased to the
  period indicated in Table \ref{table:vars_m2a}, while V57 is shown
  phased, but its intensity is given only in relative
  fluxes (in units of ten thousand ADUs). \label{fig:lc_m2}}
\end{figure*}

The variable star finding scheme explained in Section \ref{sec:isis},
apart from detecting all the variables from \citet{salinas07} within
the SAMI FOV, revealed the presence of one previously unknown variable
(V31 in Figs. \ref{fig:cmd_sam} and \ref{fig:n1261_lc}). We find a
period of 0.1 days, although this is likely a lower limit, given its
proximity to the complete time span of observations. The shape of its
light curve, together with its position in the CMD of NGC\,1261,
slightly above and redder than the upper main sequence (red symbol in
the NGC\,1261 panel in Fig. \ref{fig:cmd_sam}), suggests its
classification as an eclipsing binary, since it is too red to be
either a background delta Scuti or an RRc. As an eclisping binary, its
period is probably closer to $\sim$ 0.25 days. This variable is about
80\arcsec\, from the cluster center, so it was probably missed before
due to its faintness, and not because of crowding. Even though this
short distance to the center suggests cluster membership its nature
as a foreground object cannot be ruled out.

Finally, we point out that none of the four BSs where variability was
suspected by \citet{simunovic14} show signs of variability in our
data. The authors of that study suggested the possibility of
variability based on the fact that these four BSs had fainter F336W
magnitudes than expected, consistent with a potential variability in
the F336W magnitude. We note that our SAM photometry of these BSs is
consistent to the optical colours and magnitudes given by
\citet{simunovic14} therefore supporting their characterization in the
CMD, in particular for the proposed young collisional products as
found by stellar collision model fitting in the CMD. This rather
suggests a photometry error in their measured F336W magnitudes, or
instead a still unexplained cause for their fainter F336W magnitudes.

\subsection{M\,2}

According to the \citet{clement01} catalogue of variable stars in
Galactic globular clusters (2014 update), M\,2 hosts 42 variable
stars. Of these, 13 were discovered by the image-subtraction study of
\citet{lazaro06}, who could not transform their relative flux light
curves into magnitudes due to an anomalous psf.

We detected all the previously known variables that fall within the
SAMI FOV and further discovered 11 new RR Lyrae stars, 1 RRL/W UMa candidate (see
below) and 2 SX~Phe stars, the latter being the first SX~Phe detected in this
cluster. Table \ref{table:vars_m2a} gives positions, as well as mean
magnitudes, amplitudes and classifications. For the RRL, the poor
phase coverage prevents us from finding periods and we only give a
lower limit. Mean magnitudes and amplitudes are highly uncertain and
were derived only using the observed data and not the templates; this
implies a large scatter in the colors (see corresponding CMD in
Fig. \ref{fig:cmd_sam}), although their position in the CMD makes the classification as RRL unequivocal. The tentaive classification into RRab
or RRc is done based on the partial shape of their light curves, 
hence is also necessarily uncertain. For the SX Phe, these quantities
are measured directly from the data.

Fig. \ref{fig:lc_m2} shows the light curves of the newly discovered
variables. V57, one of the discovered SX Phe, is the only new variable
among the four clusters for which we cannot convert the relative flux
light curve into magnitudes due to blending. For V49 we measure both
the minimum and maximum light phases, showing also the characteristic
``hump'' before maximum light \citep{smith95,catelan15}. These features would
make it most likely a short-period RRc, but its low brightness puts it
close to main sequence of the cluster, instead of the HB. This is
probably an indication that this is a background variable and not a cluster
member. Alternatively, the roughly sinusoidal shape could indicate a contact
binary of the W UMa type, although the hump would remain unexplained.

\begin{figure*}
\includegraphics[width=0.98\textwidth]{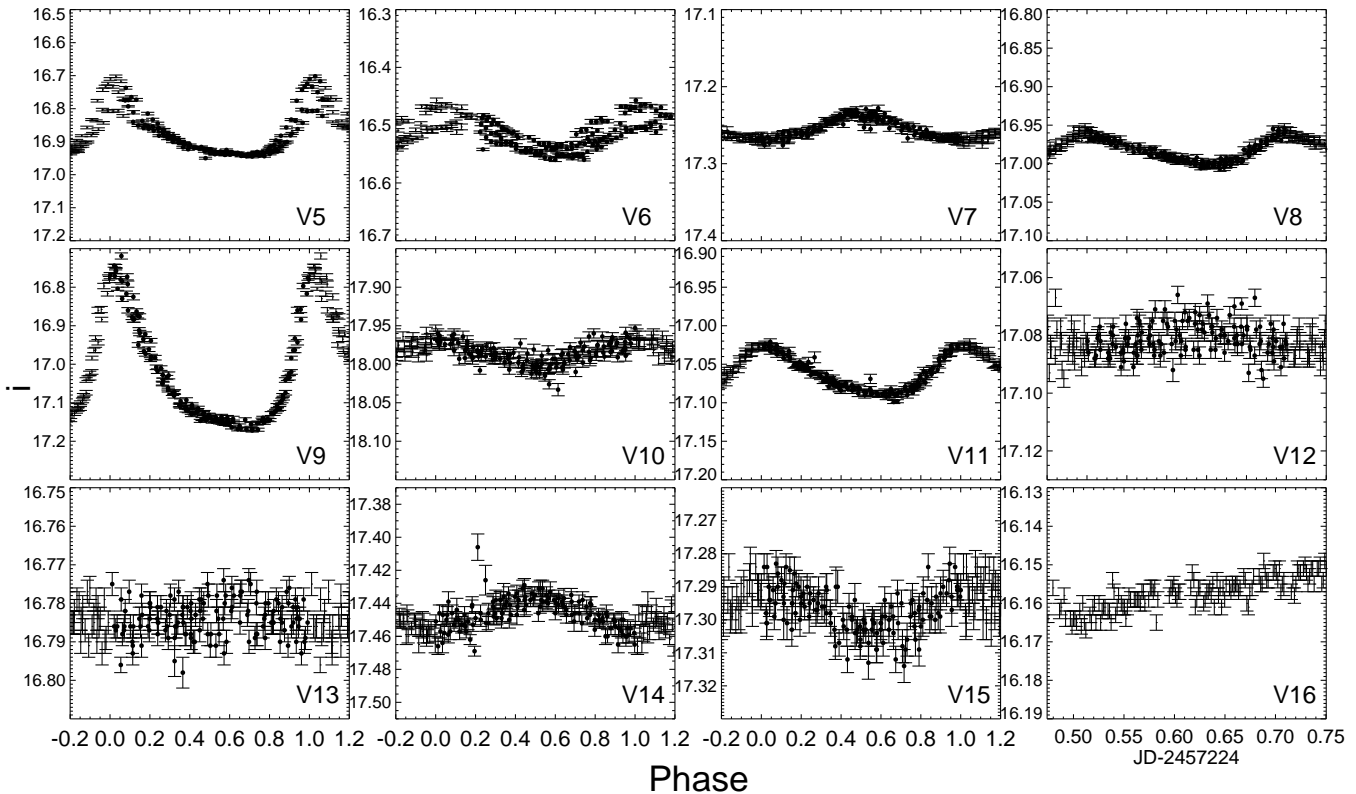}
\caption{New variables in M\,10. V5--V15 are shown phased using the
  periods given in Table \ref{table:vars_m10}, while the light curve
  of the long period variable V16 is shown in Julian dates
  only. \label{fig:lc_m10}}
\end{figure*}

Table \ref{table:vars_m2b} contains the variables discovered by
\citet{lazaro06}, all but one inside the SAMI FOV. For these we give
new coordinates, mean magnitudes and amplitudes also derived from the
template fitting. The latter two quantities were not given by
\citet{lazaro06}.

\subsection{M\,10}\label{sec:vars_m10}

The \citet{clement01} catalogue lists 4 variables in M\,10. Three of
them are long-period variables and the fourth, suspected to be an RR
Lyrae, lies outside the FOV of this study, although due to its low
metallicity and extended blue horizontal branch, few if any RRL are
expected. There is no modern CCD-based variability study of this
cluster.

Using our data, we found 12 new variables in the M\,10 field: 11 SX Phe
and a long-period red giant. Positions, periods and amplitudes for all
of them are shown in Table \ref{table:vars_m10}, while phased light
curves for the SX Phe are shown in Fig. \ref{fig:lc_m10}. Since this
cluster has the largest amount of new discoveries with completely
sampled light curves, we proceed to a more detailed description of
their characteristics.

V5 and V6 present noticeable amplitude changes which hint at pulsation
in more than one mode. V5 changes its amplitude by more than 0.1 mag
while for V6, the change is close to 0.05 mag. In both cases the time
span of observations covers only about 4 pulsation cycles, making it
impossible to find frequencies other than the dominant one.
\citet{McNamara95} suggested a broad separation between fundamental
and first-overtone pulsators based on the amplitude and shape of the
light curve; SX Phe pulsating in the fundamental mode would have
amplitudes $\Delta V \geq 0.25$ mag and asymmetrical light curves,
while first-overtone pulsators would have more sinusoidal light curves
with amplitudes $\Delta V \leq 0.20$. Using the ratio between the
amplitudes in $V$ and $I$, $A_V/A_I=1.7$ as a guide
\citep{Rodriguez07, cohen12}, we can classify V5 as a fundamental-mode
pulsator and V6 as a first-overtone pulsator.

V7 and V8 show no significant amplitude changes, but while both have
very similar amplitudes, $A_i \sim0.065$, V7 shows a sinusoidal shape,
while the light curve of V8 shows the sawtooth-like shape associated
with higher-amplitude fundamental pulsators.

\begin{figure}
\includegraphics[width=0.48\textwidth]{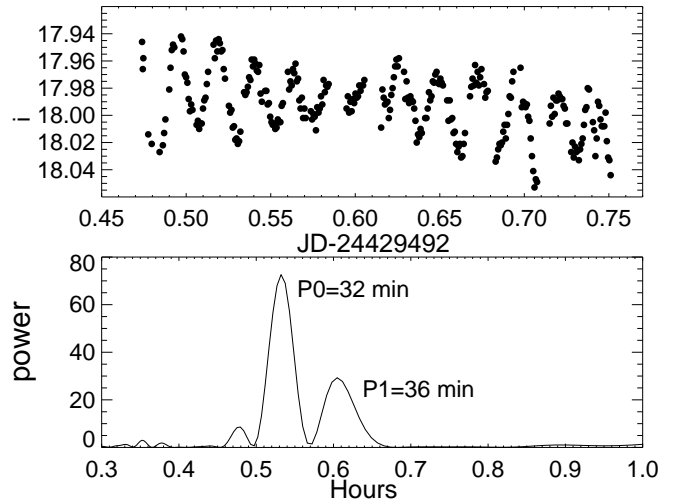}
\caption{Upper panel: light curve of V10 in M10. Lower panel: Power
  spectrum of this variable obtained with the Lomb-Scargle algorithm
  revealing the presence of two close periods in the light
  curve. \label{fig:lomb}}
\end{figure}

\begin{figure*}
\includegraphics[width=0.98\textwidth]{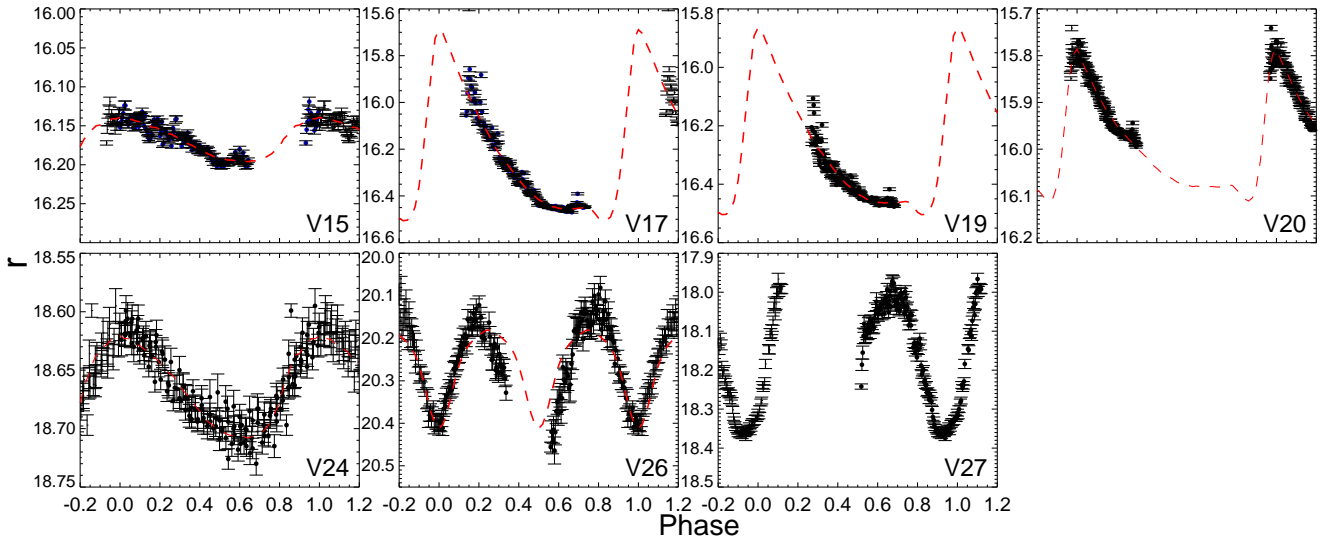}
\caption{Variables in M\,80 that had incomplete information in \citet{kopacki13}.  Dashed lines indicate the best-fit template. This is not shown for V27 where no fit was found satisfactory. \label{fig:m80_lc}}
\end{figure*}

V9 has the largest amplitude of the sample, $A_i = 0.414$, which
implies $A_V\sim0.7$. Its amplitude experiences a change of about 0.1
mag in 5 pulsation cycles, but the Lomb-Scargle exploration does not
reveal the presence of another period. From the catalogue of SX Phe in
globular clusters by \citet{cohen12} only a handful of SX Phe have
larger amplitudes. Large amplitudes are more common among the
high-metallicity counterparts of SX Phes, the $\delta$ Scuti stars
\citep[e.g.][]{vivas13}, particularly in the subclass of
high-amplitude $\delta$~Scutis \citep[][and references
  therein]{catelan15}. Its very blue color (see
Fig. \ref{fig:cmd_sam}), consistent with a star belonging to the
extended HB instead of the blue straggler region, is likely an effect
of blending in the $g$ band images which have significantly poorer
quality compared to the $i$ images. In the \textit{HST} photometry it
appears brighter than the BSs of similar color, which could signal it
as a foreground object, although its distance of $\sim25\arcsec$ from
the cluster center would rather give an indication of cluster
membership.

V10 presents some interesting features which make its final
classification also uncertain. First, it has a very short main period
of 0.022 days. Only two known SX Phe in $\omega$ Cen have slightly
shorter periods \citep{Kaluzny04}. Moreover, the Lomb-Scargle analysis
reveals the presence of another period of 0.025 days (see
Fig. \ref{fig:lomb}). It is fainter than the rest of the SX Phe in
M\,10 following the expectation from the period-luminosity relation in
SX Phe \citep[e.g.][]{McNamara95}, but it is also bluer than the rest
except V9. Like in the case of V9, the bluer color could be explained
by partial blending in the $g$ band. The presence of two periods would
be associated with two pulsation modes. An alternative explanation
could be a HW Vir-type variable, that is, a subdwarf B (sdB) star with
a faint eclipsing companion \citep{menzies86}, but the lack of strong
flux in the F275W band from the \textit{HST} photometry disfavors the
presence of a very hot star.

V11 has no remarkable features other than its sawtooth-like shape,
which probably makes it a pulsator in the fundamental mode.


V12 to V15 are unremarkable SX Phe stars with mostly sinusoidal light
curves and low amplitudes ($\sim$ 0.02 mag) that appear very noisy in
the SAM data. They appear well within the BS region in the SAM CMD.

Finally, V16 is located in the lower part of the RGB (see
Fig. \ref{fig:cmd_sam}), has a monotonic increase in luminosity with
an amplitude $A_i = 0.02$ during the time of observations.  We
classify it as a long-period variable, a common occurrence among RGB
stars, , though its exact nature cannot be established based solely on
the present data.

\subsection{M\,80}\label{sec:vars_m80}

The \citet{clement01} catalogue lists 33 variables in M\,80. It was
also the host of the only known classical nova in a Galactic GC with
convincing evidence, an event which took place in 1860
\citep{luther60}, which became known accordingly as Nova 1860 AD or
T~Scorpii.

Variable star searches in the cluster after the latest Clement
catalogue update of 2010 were conducted by \citet{kopacki13} and
\citet{figuera15}. While \citet{kopacki13} found 9 new RRL, 4 SX Phe,
two eclipsing binaries and two other periodic variables of unknown
type, the work of \citet{figuera15} found only six new long-period
variables.

Our search recovered most of the short-period variables of the
previous studies within the SAMI FOV with the exception of V33, an SX
Phe discovered by \citet{thomson10} very close to the cluster center
which was neither recovered by \citet{kopacki13}, but we did not find
any previously unknown variable.

Even though the short time span of our observations does not help in
clearing the nature of the variables classified as ``unknown'' by
\citet{kopacki13} given their periods larger than two days, we give
new colors and magnitudes for the variables V15, V17, V19, V20, V24,
V26 and V27, that were not measured by \citet{kopacki13} due to
difficulties in transforming relative fluxes into magnitudes. The new
magnitudes, amplitudes and colors for these variables can be seen in
Table \ref{table:vars_m80}, while light curves can be seen in
Fig. \ref{fig:m80_lc}. The $g$ magnitude for the variables is simpy
taken as the average of the magnitudes in the three $g$ frames,
therefore the position of the variables in the CMD
(Fig. \ref{fig:cmd_sam}) is very tentative. Finally, our limited
observations did not reveal signs of variability either in the
vicinity of T~Sco or of the other cataclismic variable candidates
reported by \citet{shara05} and \citet{dieball10}.

\section{M\,80: A blue straggler-rich cluster?} \label{sec:m80}

\begin{figure*}
\includegraphics[width=0.98\textwidth]{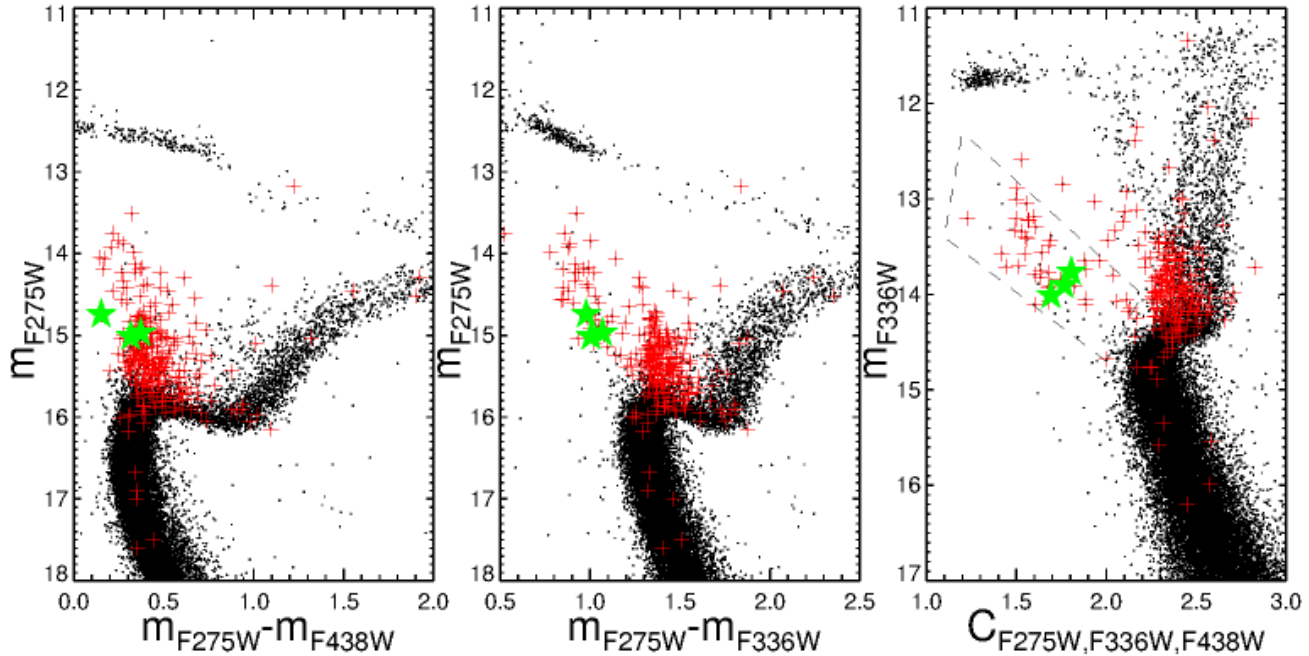}
\caption{CMD of the blue straggler region in M80 using
  \textit{HST}/WFC3/UVIS filters F275W, F336W and F438W. BS candidates
  from \citet{ferraro99} are shown in red crosses. Left panel: the
  F275W--F438W CMD. Central panel: the F275W--F336W CMD.  Right panel:
  the color index $C_{\rm F275W,F336W,F438W}$ gives the best
  separation for the genuine BSs. The dashed lines define the BS
  region considered in our study. The three green stars indicate the
  positions of the SX Phe inside the UVIS FOV.
\label{fig:m80}}
\end{figure*}

\citet{ferraro99} studied the content of blue stragglers in M\,80 using
the \textit{HST}/WFPC2 filters F225W and F336W. Ultraviolet
observations in globular clusters are helpful to study hot stars such
as blue stragglers and horizontal branch stars given the lower flux
that the otherwise dominant RGB population has in these bandpasses.  Based on a
color and magnitude selection they found a very high number of 305 BSs
within the WFPC2 FOV, that would make it the GC with the highest
number of BSs known in the Galaxy.

\citet{ferraro99} argued that stellar density, even though very high
in the center of this cluster, could not explain by itself this
overpopulation of BSs, but instead that the process leading to core
collapse enhances the formation of binaries \citep[e.g.][]{meylan97}
that goes in hand with an increase in the number of BSs.  

This view was challenged by \citet{dieball10} who, using
\textit{HST}/ACS FUV and NUV filters found only 75 BSs, the same
number as in M 15, making M\,80 unremarkable regarding its BSs content.

Using newly available \textit{HST}/UVIS data of M80 obtained in the
filters F275W, F336W and F436W, we reassess the BS content of
M80. Using the UVIS photometry from Section \ref{sec:hst} we constructed
color-magnitude diagrams, cross-matching the positions of the BS
candidates from \citet{ferraro99} to our photometry using
\textsc{CataXcorr/CataComb}. BS candidates from \citet{ferraro99} are
shown in Fig. \ref{fig:m80} as red crosses.

The different panels in Fig. \ref{fig:m80} show how the blue straggler
region is populated differently depending on the bandpasses used. In
the F275W--F436W color, most of the candidates from \citet{ferraro99}
fall inside the BS region (left panel), but using the bluer color
F275W--F336W already reveals a separation between blue stagglers and a
population of fainter ``yellow stragglers'' \citep{hesser84}, with
colors slightly redder than the MSTO (middle panel). Finally, using
the pseudo-color $C_{\rm F275W,F336W,F438W}=(m_{\rm F275W}-m_{\rm
  F336W})-(m_{\rm F336W}-m_{\rm F438W})$, which has been shown to be
very effective to separate multiple stellar populations
\citep{milone13,piotto15}, the separation between the real blue
stragglers and this yellow population is clearly established.

``Yellow stragglers'' can be considered as the evolution of 
BSs \citep[e.g.][]{landsman97}, but in this case the very 
narrow color range they occupy, considered in tandem with the very 
high central density, points rather to blends due to crowding, despite 
the resolving power of \textit{HST}.

Using the $C_{\rm F275W,F336W,F438W}$ color index we make a new
determination of the number of BSs in M80. We consider as BSs the
stars within the bounding box shown in Fig. \ref{fig:m80} (right
panel) counting up to 79 stars in good agreement with
\citet{dieball10}. The significance of this number can be assessed by
comparing to a control population within the cluster. In particular,
following \citet{ferraro99}, we define the specific frequency of BSs
as the ratio to horizontal branch stars, $F^{\rm BSS}_{\rm HB}=N_{\rm
  BSS}/N_{\rm HB}=0.22$. If we consider the specfic frequencies
derived for 56 GCs by \citet{piotto04}, clusters with luminosities
similar to M80 have a range of $0.15\lesssim F^{\rm BSS}_{\rm
  HB}\lesssim0.4$, indicating that the specific frequency in M\,80 is
normal.  Furthermore, our variability study (Section
\ref{sec:vars_m80}) and the previous variability studies
\citep{kopacki13,figuera15} do not reveal the presence of a large
amount of binaries (including eclipsing binaries) that should also be
enhanced during a state prior to core collapse
\citep[e.g.][]{meylan97} and that should be readily detectable. We
therefore conclude that M\,80 does not host an unusually large number
of BSs and may not be close to core collapse, in contrast to the early
claim of \citet{ferraro99}.

\section{Summary and conclusions} \label{sec:conclusions}

In this paper we presented time-series photometry of four globular
clusters: M\,2, M\,10, M\,80 and NGC 1261.  These images were obtained
using SAM, the adaptive optics module at the SOAR Telescope. The
correction of the ground-layer turbulence done by SAM provides sharper
images, which allow absolute photometry closer to the centers of
globular clusters than non-AO ground-based observations, and in this
way a larger number of light curves obtained through image subtraction
can be calibrated to magnitudes.

We studied the four clusters using the image subtraction technique
\citep{alard98}, finding in total 28 new variables. In M\,2, we
discovered 12 new RRL stars and 3 SX~Phe stars. In M\,10, we found
11 SX Phe stars and one long-period variable, while in NGC\,1261 we
found an eclipsing W UMa-type variable. The light curves of all but
one of these new variables were transformed into magnitudes.

We also reviewed the claim that M\,80 is the GC with the highest
number of BSs in our Galaxy. Based on \textit{HST}/UVIS data we found
$\sim$ 80 BSs, significantly less than the 305 announced by
\citet{ferraro99}, suggesting that the cluster may have a lower
specific fraction of BSs than previously suspected.

In summary, we have shown that ground-layer AO-assisted imaging has
the potential to pierce deeper into crowded environments than normal
ground-based observations, being very helpful for the detection and
calibration of variable stars. Even though the image quality delivered by SAM
cannot rival the power of \textit{HST}, the possibility to use it for
longer periods of time makes it a very good match to variability and
monitoring programs in crowded environments. Furthermore, the
performance of image subtraction against high surface brightness
backgrounds (e.g. within distant galaxies) has been shown to be
greatly improved when sharper image quality is used
\citep[e.g.][]{rau08,kerins10}.

\acknowledgments

We thank the anonymous referee for a fast report that helped improve
the presentation of our results. We thank Andrei Tokovinin and C\'esar
Brice\~ no for their assistance during the SOAR/SAM runs. JS
acknowledges partial support from NSF grant AST-1308124 and the
Packard Foundation. Support for M.C. and R.C.R. is provided by the
Ministry for the Economy, Development, and Tourism's Millennium
Science Initiative through grant IC\,120009, awarded to the Millennium
Institute of Astrophysics (MAS). M.C. acknowledges additional support
by Proyecto Basal PFB-06/2007 and by FONDECYT grant
\#1141141. Supported by the Gemini Observatory, which is operated by
the Association of Universities for Research in Astronomy, Inc., on
behalf of the international Gemini partnership of Argentina, Brazil,
Canada, Chile, and the United States of America.  Based on
observations made with the NASA/ESA Hubble Space Telescope, obtained
from the data archive at the Space Telescope Science Institute. STScI
is operated by the Association of Universities for Research in
Astronomy, Inc. under NASA contract NAS 5-26555.

\textit{Facility: }\facility{SOAR}

\bibliography{salinas_apj}

\begin{deluxetable*}{lccllll}
\tablewidth{0pt}
\tablecaption{Variables in NGC 1261}
\tablehead{
\colhead{ID} & \colhead{RA (J2000)} & \colhead{Dec (J2000)}& \colhead{$\langle r\rangle$} & \colhead{$A_r$} &  \colhead{$P$(d)} & \colhead{Type}
}
\startdata
V22 & 03 12 16.49 & --55 13 38.10 & 17.00: & 0.21: & 0.302  & RRc\\
V24 & 03 12 14.43 & --55 13 34.77 & 16.97: & 0.83: & 0.626  & RRab\\
V26 & 03 12 17.05 & --55 12 43.92 & 18.66  & 0.12  & 0.0799 & SXPhe\\
V27 & 03 12 14.65 & --55 13 06.50 & 16.73: & 0.11: & 0.341  & RRc\\
V28 & 03 12 13.53 & --55 13 00.80 & 17.00: & 0.15: & 0.287  & RRc \\
V29 & 03 12 13.05 & --55 13 20.45 & 16.98: & 0.15: & 0.593  & RRab\\
V30 & 03 12 16.58 & --55 12 53.96 & 18.42  & 0.07  & 0.0591 & SXPhe\\
V31 & 03 12 18.70 & --55 14 16.00 & 20.59  & 0.16  & $>0.1$   & W UMa?
\enddata
\label{table:vars_n1261}
\tablecomments{The table contains all the variables discovered by
  \citet{salinas07} in the SAMI FOV that were not saturated, plus the
  newly found V31. For the known variables, periods and classification
  are taken from \citet{salinas07}. Colons indicate uncertain values.}
\end{deluxetable*}

\begin{deluxetable*}{lcclccl}
\tablewidth{0pt}
\tablecaption{New variables in M\,2}
\tablehead{
\colhead{ID} & \colhead{RA (J2000)} & \colhead{Dec (J2000)}& \colhead{$\langle i\rangle$} & \colhead{$A_i$} &  \colhead{$P$(d)} & \colhead{Type}
}
\startdata
V43 & 21 33 26.44 & --00 49 29.3 & 16.00: &$>0.12$ & $>0.12$ & RRab?\\
V44 & 21 33 26.69 & --00 49 21.8 & 15.81: &$>0.05$ & $>0.12$ & RRc?\\
V45 & 21 33 26.13 & --00 49 22.2 & 15.91: &$>0.09$ & $>0.12$ & RRab?\\
V46 & 21 33 27.45 & --00 49 15.5 & 15.94: &$>0.05$ & $>0.12$ & RRab?\\
V47 & 21 33 27.40 & --00 49 05.4 & 15.91: &$>0.09$ & $>0.12$ & RRc?\\
V48 & 21 33 27.51 & --00 49 07.5 & 15.76: &$>0.10$ & $>0.12$ & RRab?\\
V49 & 21 33 30.45 & --00 50 29.6 & 18.82: & $\sim 0.20$ &  $>0.12$ &RRc?/W UMa?\\
V50 & 21 33 26.63 & --00 49 11.2 & 16.01: &$>0.03$ & $>0.12$ & RRc?\\
V51 & 21 33 24.94 & --00 48 43.3 & 15.68: &$>0.14$& $>0.12$ & RRab?\\
V52 & 21 33 25.12 & --00 49 24.4 & 16.12: & $\sim0.26$& $>0.12$ & RRc?\\
V53 & 21 33 28.24 & --00 49 35.4 & 16.00: & $>0.24$& $>0.12$ & RRc?\\
V54 & 21 33 27.55 & --00 49 29.1 & 16.11: & $>0.36$ & $>0.12$ & RRab?\\
V55 & 21 33 26.37 & --00 49 18.1 & 15.54: & $>0.20$ & $>0.12$ & RRab?\\
V56 & 21 33 31.62 & --00 50 13.0 & 18.11 & 0.10 & 0.0468  &SX Phe\\
V57 & 21 33 27.54 & --00 49 21.4 & -- & -- & 0.0686 & SX Phe

\enddata
\label{table:vars_m2a}
\tablecomments{The table contains all the new variables discovered in
  M\,2. Colons indicate uncertain values.}
\end{deluxetable*}

\begin{deluxetable*}{lcccccl}
\tablewidth{0pt}
\tablecaption{Variables in M\,2 from \citet{lazaro06}}
\tablehead{
\colhead{ID} & \colhead{RA (J2000)} & \colhead{Dec (J2000)}& \colhead{$\langle i\rangle$} & \colhead{$A_i$} &  \colhead{$P$(d)} & \colhead{Type}
}
\startdata
V36 & 21 33 30.71 & -00 49 13.5 & 16.21: & 0.08: & 0.27078 & RRc\\
V37 & 21 33 26.04 & -00 49 18.1 & 15.79: & 0.19: & 0.56668 & RRab\\
V38 & 21 33 31.20 & -00 49 23.8 & 15.84: & 0.23: & 0.80735 & RRab\\
V39 & 21 33 27.38 & -00 50 07.2 & 16.08: & 0.13: & 0.60781 & RRab\\
V40 & 21 33 25.66 & -00 49 16.3 & 16.11: & 0.18: & 0.75173 & RRab\\
V41 & 21 33 28.02 & -00 49 24.3 & 15.68: & 0.32: & 0.60532 & RRab\\
V42 & 21 33 28.42 & -00 49 54.6 & 16.09: & 0.27: & 0.32801 & RRc
\enddata
\label{table:vars_m2b}
\tablecomments{Periods and classification are from \citet{lazaro06},
  while positions, mean magnitudes and amplitudes come from this work.
  Colons indicate uncertain values.}
\end{deluxetable*}

\begin{deluxetable*}{lcccccl}
\tablewidth{0pt}
\tablecaption{New variables in M\,10}
\tablehead{
\colhead{ID} & \colhead{RA (J2000)} & \colhead{Dec (J2000)}& \colhead{$\langle 
i\rangle$} & \colhead{$A_i$} &  \colhead{$P$(d)} & \colhead{Type}
}
\startdata
V5  & 16 57 08.59 & --04 06 16.3 & 16.875 & 0.267 & 0.058 & SX Phe\\
V6  & 16 57 10.71 & --04 05 33.3 & 16.499 & 0.115 & 0.060 & SX Phe\\
V7  & 16 57 10.35 & --04 07 03.0 & 17.251 & 0.051 & 0.048 & SX Phe\\
V8  & 16 57 08.29 & --04 05 10.0 & 16.978 & 0.055 & 0.051 & SX Phe\\
V9  & 16 57 10.65 & --04 05 50.8 & 17.040 & 0.414 & 0.051 & SX Phe/ $\delta$ Sct?\\
V10 & 16 57 08.43 & --04 06 54.7 & 17.977 & 0.086 & 0.022 & SX Phe/HW Vir?\\
V11 & 16 57 10.81 & --04 05 55.9 & 17.064 & 0.078 & 0.048 & SX Phe\\
V12 & 16 57 04.04 & --04 06 06.9 & 17.091 & 0.020 & 0.023 & SX Phe\\
V13 & 16 57 08.80 & --04 06 24.5 & 16.775 & 0.016 & 0.036 & SX Phe\\
V14 & 16 57 09.20 & --04 06 05.3 & 17.449 & 0.066 & 0.038 & SX Phe\\
V15 & 16 57 13.28 & --04 05 48.7 & 17.297 & 0.031 & 0.035 & SX Phe\\  
V16 & 16 57 06.21 & --04 06 42.2 & 16.16: & 0.02: & $>0.3$ & LPV?  
\enddata
\label{table:vars_m10}
\tablecomments{The table contains all the new variables discovered in
  M\,10. Colons indicate uncertain values.}
\end{deluxetable*}

\begin{deluxetable*}{lccllll}
\tablewidth{0pt}
\tablecaption{Variables in M\,80}
\tablehead{
\colhead{ID} & \colhead{RA (J200)} &\colhead{Dec (J2000)} & \colhead{$\langle r\rangle$} & 
\colhead{$A_r$}& \colhead{$P$(d)} & \colhead{Type} 
}

\startdata
V15&16 17 04.04 & --22 58 40.2 &  16.17: & 0.08: & 0.3479  &RRc\\
V17&16 17 04.61 & --22 58 40.0 &  16.20: & 0.82: & 0.4154  &RRc\\
V19&	16 17 02.11 & --22 58 29.5 &  16.28: & 0.64: & 0.5956  &RRab\\
V20&16 17 03.26 & --22 58 37.5 &  16.00: & 0.32: & 0.7448  &RRab\\
V24&16 17 02.99 & --22 59 21.1 &  18.67  & 0.13  & 0.04941&SX Phe\\
V26&	16 17 04.49 & --22 59 17.9 &  20.27: & 0.38: & 0.3190  &W UMa\\
V27&	16 17 04.02 & --22 58 26.5 &  18.13: & 0.41: & 0.4117  & W UMa
\enddata
\label{table:vars_m80}
\tablecomments{The tables contains variables from \citet{kopacki13} that had incomplete information. Position, classification and periods come from \citet{kopacki13}, while mean magnitudes and amplitudes are from the present study. Colons indicate uncertain values.}
\end{deluxetable*}

\end{document}